\documentclass[11pt,english]{article}
\usepackage[T1]{fontenc}
\usepackage[latin1]{inputenc}
\usepackage{amsmath}
\usepackage{amssymb}

\makeatletter
 \usepackage{verbatim}

\usepackage{a4wide}

\usepackage{babel}
\makeatother
\begin{document}
\setcounter{page}{0}

$\,\,$
\vspace{1cm}

\begin{center}\textbf{\LARGE Hitchin's equations and integrability }\end{center}{\LARGE \par}

\begin{center}\textbf{\LARGE of BPS $Z_{N}$ strings in Yang-Mills
theories }\end{center}{\LARGE \par}

\begin{center}

\vspace{1cm}

{{\bf Marco A. C. Kneipp}} 

\end{center}

\begin{center}{\em Universidade Federal de Santa Catarina (UFSC)%
\footnote{E-mail address: kneipp@fsc.ufsc.br.%
},\\ 

Departamento de F\'\i sica, CFM,\\

Campus Universit\'ario, Trindade,\\

88040-900, Florian\'opols, Brazil.  \\ }

\end{center}

\vspace{0.3cm}

\begin{abstract}
We show that $Z_{N}$ string's BPS equations are equivalent to the
Hitchin's equations (or self-duality equation) and also to the zero
curvature condition. We construct a general form for BPS $Z_{N}$
string solutions for arbitrary simple gauge groups with non-trivial
center. Depending on the vacuum solutions considered, the $Z_{N}$
string's BPS equations reduce to different two dimensional integrable
field equations. For a particular vacuum we obtain the equation of
affine Toda field theory. 

\vfill Keywords: Integrable Equations in Physics, Integrable Hierarchies.

\thispagestyle{empty}
\end{abstract}
\newpage

\section{Introduction}

In $SU(N)$ QCD, it is believed that particle confinement in the strong
coupling regime happens due to chromoelectric strings (QCD strings).
Many properties of the QCD strings have been studied intensely in
the last years using lattice calculation. On the other hand, it is
believed that QCD strings in the strong coupling may be dual to chromomagnetic
strings in the Higgs phase in weak coupling, which are easier to study
analytically. Since QCD strings in confining phase should be formed
only by $SU(3)$ gauge fields and not $U(1)$ gauge fields, in recent
years we are analyzing some properties of chromomagnetic $Z_{N}$
strings solutions which appear in a theory with non-Abelian simple
gauge group $G$ (without $U(1)$ factors) broken to its center.

The $Z_{N}$ string solutions have many features similar to the QCD
strings. In particular they are associated to coweights of representations
of $G$ (or equivalently to weights of the dual group%
\footnote{We shall consider the dual group $G^{\vee}$ as the covering group
associated to the dual algebra $\mathfrak{g}^{\vee}$.%
} $G^{\vee}$) and their topological sectors are associated to the
center elements of the gauge group $G$. More precisely, the coweights
of $G$ can be separated in cosets associated to nodes of the extended
Dynkin diagram of $G$. All $Z_{N}$ string solutions associated to
coweights in a given coset belong to the same topological sector \cite{k2003}.
The $Z_{N}$ strings associated to the fundamental weights of different
representations can have different tensions and for different vacuum
solutions, the BPS bounds for the tensions can satisfying either the
sine law scaling or the Casimir scaling \cite{k2003}\cite{kneippsine},
differently from the non-Abelian semi-local BPS string solutions with
gauge group $SU(N)\times U(1)$ where the tension is only due magnetic
flux in the $U(1)$ direction and it depends on the $U(1)$ winding
number\cite{G x U(1)}. It is important to note that the Casimir scaling
and the sine law scaling considered in \cite{k2003}\cite{kneippsine}
are lower bounds for the non-BPS $Z_{N}$ string tensions. 

Previously we analyzed the $Z_{N}$ string in soft broken ${\cal N}=2$
\cite{kbrockill2001}\cite{k2002} and ${\cal N}=4$ \cite{k2003}
Super Yang-Mills theories, but in \cite{kneippsine} and here we do
not constraint the potential to be supersymmetric since we are interested
in studying some general properties at the classical level of the
$Z_{N}$ strings which may be useful for QCD and not necessarily confinement
in supersymmetric theories. The $Z_{N}$ strings does not necessarily
point in a direction in the Cartan subalgebra (CSA). However, since
the monopoles' magnetic flux is in the CSA \cite{gno}, we only consider
$Z_{N}$ string solutions with flux in the CSA which are the relevant
for confinement of these monopoles \cite{k2002}\cite{k2003} which
may be dual to particle confinement. This result is analogous to the
Abelian dominance observed in QCD.

In the present work we show that $Z_{N}$ string's BPS equations \cite{kbrockill2001}\cite{k2003}
for Yang-Mills theories with scalars in the adjoint are equivalent
to the Hitchin's equations \cite{hitchinselfduality} and consequently
to the four dimensional self-duality equation \cite{hitchinselfduality}
and also equivalent to the zero curvature condition\cite{bonora1},
implying that this set of BPS $Z_{N}$ string solutions in non-Abelian
Yang-Mills theories is (quasi-)integrable. Integrability of BPS vortices
in Abelian-Higgs theory was recently considered in \cite{integrabilityabelianHiggs}.
Integrability of other soliton solutions for theories in dimensions
higher than two are analyzed in \cite{luiz}. In recent years, integrability
also had a renewed interest in gauge and string theories \cite{zarembo}.
On the other hand, Hitchin's equations appeared in many distinct problems
as for example in Matrix string theory \cite{bonora1}\cite{matrix}
and more recently in connection with the geometric Langlands program
\cite{langland}. Integrability of Hitchin's equations and self-duality
equations are also discussed in \cite{Hitchin-Ward}\cite{ward}.

The equivalence of BPS $Z_{N}$ string equations with Hitchin's equations,
self-duality equations and zero curvature condition is interesting
because allows us to apply methods and results of these systems to
$Z_{N}$ string solutions and vice-versa. 

In the $U(1)$ Abelian-Higgs theory, the fields of rotationally symmetric
BPS string solutions can be written as functions of a field which
satisfies Liouville's equation plus a constant and a singularity at
the origin \cite{taubes}\cite{rebbi}. In this work we generalize
this result for the $Z_{N}$ strings showing that, for a particular
vacuum responsible for the gauge symmetry breaking, the fields of
the rotationally symmetric BPS Z(N) string solutions are functions
of a field which satisfies affine Toda field equation with a singularity
at the origin.

In this paper we introduce, in section 2, some general results for
BPS $Z_{N}$ strings and show the equivalence of the $Z_{N}$ string
BPS conditions with the Hitchin's equations, and consequently to self-duality
equation and the zero curvature condition. In section 3 we construct
an Ansatz for the $Z_{N}$ strings and show that the $Z_{N}$ string's
BPS equations reduce to two dimensional integrable theories equations.
In section 4 we show that for a particular vacuum, the BPS $Z_{N}$
string solutions reduces to the equation of affine Toda theory which
is a deformation of conformal Toda theory. In section 5 we analyze
the special case of rotationally symmetric solutions. These solutions
resemble the Riemannian or stringy instantons of Matrix string theories\cite{bonora1}\cite{matrix}.

\section{BPS $Z_{N}$ strings equations and the Hitchin's equations}

Let us consider Yang-Mills-Higgs theories with arbitrary gauge group
$G$ which is simple, connected and simply connected. In order to
exist strings and confined monopoles we shall consider theories with
two complex scalars fields $\phi_{s},\, s=1,2$, in the adjoint representation
of $G$. We also consider that the vacuum solutions $\phi_{1}^{\textrm{vac}}$,
$\phi_{2}^{\textrm{vac}}$ produce the symmetry breaking pattern\begin{equation}
G\,\stackrel{\phi_{1}^{{\scriptstyle \textrm{vac}}}}{\rightarrow}\, U(1)^{r}\,\stackrel{\phi_{2}^{{\scriptstyle \textrm{vac}}}}{\rightarrow}C_{G},\label{1}\end{equation}
 where $r$ is the rank of $G$ and $C_{G}$ its center, which we
consider to be nontrivial. The Lagrangian of the theory we are to
study is\begin{equation}
\mathcal{L}=-\frac{1}{4}G_{a\mu\nu}G_{a}^{\mu\nu}+\sum_{s=1}^{2}\frac{1}{2}\left(D_{\mu}\phi_{s}\right)_{a}^{*}\left(D^{\mu}\phi_{s}\right)_{a}-V(\phi,\phi^{*})\label{4}\end{equation}
where $a$ is a Lie algebra index, $D_{\mu}=\partial_{\mu}+ie[W_{\mu},$~~{]}
and $G_{\mu\nu}=\partial_{\mu}W_{\nu}-\partial_{\nu}W_{\mu}+ie[W_{\mu},W_{\nu}]$.
Let\begin{eqnarray}
z & = & x^{1}+ix^{2},\nonumber \\
\partial_{z} & = & \frac{1}{2}\left(\partial_{1}-i\partial_{2}\right),\label{44}\\
W_{z} & = & \frac{1}{2}\left(W_{1}-iW_{2}\right),\nonumber \end{eqnarray}
 and $B_{i}=-\epsilon_{ijk}G_{jk}/2$. For a static string solution
with cylindrical symmetry in the $x_{3}$ direction, the string BPS
equations for a theory with gauge group $G$ without $U(1)$ factors
are\cite{kbrockill2001}\cite{k2003}%
\footnote{For the sake of simplicity we shall only consider the string solutions
which have positive flux $\Phi_{\textrm{st}}$. For the antistrings,
one must use opposite signs in some of these equations as discussed
in our previous works.%
}\begin{eqnarray}
B_{3a} & = & -d_{a},\label{5a}\\
D_{z}\phi_{s} & = & 0,\label{5b}\\
D_{\bar{z}}\phi_{s}^{\dagger} & = & 0,\label{5e}\\
V(\phi,\phi^{*})-\frac{1}{2}\left(d_{a}\right)^{2} & = & 0,\label{5c}\end{eqnarray}
 with \[
d=\frac{e}{2}\left\{ \sum_{s=1}^{2}\left[\phi_{s}^{\dagger},\phi_{s}\right]-mRe(\phi_{1})\right\} ,\]
 where $m$ is a non-negative mass parameter. The string tension satisfies
the bound\begin{equation}
T\geq\frac{me}{2}\left|\phi_{1}^{\textrm{vac}}\right|\left|\Phi_{\textrm{st}}\right|\label{7a}\end{equation}
where \begin{equation}
\Phi_{\textrm{st}}=\frac{1}{\left|\phi_{1}^{\textrm{vac}}\right|}\int d^{2}x\textrm{Tr}\left[Re\left(\phi_{1}\right)B_{3}\right]\label{7b}\end{equation}
is the string flux in the $\phi_{1}$ direction, with the integral
being taking in the plane orthogonal to the string. The equality in
Eq. (\ref{7a}) happens only for the strings satisfying the BPS equations.
We shall consider \begin{equation}
V(\phi,\phi^{*})=\frac{1}{2}\left(d_{a}\right)^{2},\label{8}\end{equation}
which guarantee that equation (\ref{5c}) is automatically fulfilled.
Note that the BPS equation (\ref{5c}) does not restrict the potential
to have this form. In \cite{kbrockill2001}\cite{k2002}\cite{k2003}
we considered soft broken ${\cal N}=2$ and ${\cal N}=4$ potentials.
Similarly to the Prasad-Sommerfield limit \cite{prasad-sommerfield}
for BPS monopoles, we take the limit $m\rightarrow0$ \cite{kbrockill2001}
in order for the BPS string equations to be consistent with the equations
of motion and retain the potential terms responsible for the breaking
of $G$ into its center $C_{G}$. Note that in this limit, the tensions
$T\rightarrow0$, although the ratio of the string tensions are finite
and can satisfy the Casimir scaling law or the sine law as discussed
in\cite{kneippsine} and in section 4. 

Let $g$ be the Lie algebra associated to the gauge group $G$ and
let the generators $H_{i},\, i=1,\,2,\,...,\, r$, form a basis for
the Cartan subalgebra (CSA) $h$ with rank $r$. Let us adopt the
Cartan-Weyl basis in which\begin{eqnarray*}
\textrm{Tr}\left(H_{i}H_{j}\right) & = & \delta_{ij},\\
\textrm{Tr}\left(E_{\alpha}E_{\beta}\right) & = & \frac{2}{\alpha^{2}}\delta_{\alpha+\beta},\end{eqnarray*}
 where the trace is taken in the adjoint representation. In this basis,
the commutation relations read \begin{eqnarray}
\left[H_{i},E_{\alpha}\right] & = & \left(\alpha\right)^{i}E_{\alpha},\label{3.1}\\
\left[E_{\alpha},E_{-\alpha}\right] & = & \frac{2\alpha}{\alpha^{2}}\cdot H,\nonumber \end{eqnarray}
where $\alpha$ are roots and the upper index in $\left(\alpha\right)^{i}$
means the component $i$ of $\alpha$. We denote by $\alpha_{i}$
and $\lambda_{i}$, $i=1,2,...,\, r$ , the simple roots and fundamental
weights of $g$ respectively and \begin{equation}
\alpha_{i}^{\vee}=\frac{2\alpha_{i}}{\alpha_{i}^{2}},\,\,\,\,\,\lambda_{i}^{\vee}=\frac{2\lambda_{i}}{\alpha_{i}^{2}}\label{3.1a}\end{equation}
are the simple co-roots and fundamental co-weights of $g$, and are
also the simple roots and fundamental weights of the dual algebra
$g^{\vee}$. They satisfy the relations\[
\alpha_{i}\cdot\lambda_{j}^{\vee}=\alpha_{i}^{\vee}\cdot\lambda_{j}=\delta_{ij}.\]
 Moreover, \begin{equation}
\alpha_{i}=K_{ij}\lambda_{j}\label{3.3}\end{equation}
where \begin{equation}
K_{ij}=\frac{2\alpha_{i}\cdot\alpha_{j}}{\alpha_{j}^{2}}\label{3.4}\end{equation}
is the Cartan matrix associated to $g$. We denote by $\psi$ the
highest root of $g$. Considering the convention that $\psi^{2}=2$,
the highest root can be written as\begin{equation}
\psi=\sum_{i=1}^{r}m_{i}\alpha_{i}^{\vee}\label{3.5}\end{equation}
 where $m_{i}$ are integers which are the levels (or marks) of the
fundamental representations which have $\lambda_{i}$ as highest weights.
For $SU(n)$, $m_{i}=1$ for $i=1,\,2,\,...,\, n-1$.

In order to produce the symmetry breaking (\ref{1}) we can consider
a general vacuum solution\begin{eqnarray}
\phi_{1}^{\textrm{vac}} & = & v\cdot H\,,\,\,\,\,\,\,\,\,\,\,\,\, v=v_{i}\lambda_{i}^{\vee},\label{3.18}\\
\phi_{2}^{\textrm{vac}} & = & \sum_{l=0}^{r}b_{l}E_{-\alpha_{l}},\label{3.18b}\end{eqnarray}
where $\alpha_{0}=-\psi$ is the negative of the highest weight, $v_{i}$
are non-vanishing real constants, $b_{l}$ are real constants and
for \textbf{$l=1,2,\,...,\, r$} they can not vanish in order to $G$
to be broken into $C_{G}$. Comparing with the general vacuum solutions
considered in \cite{k2003}\cite{kneippsine}, in (\ref{3.18b}) we
add a term associated to the root $-\alpha_{0}$ which does not change
the symmetry breaking but change some properties of the vacuum solutions
as we shall explain bellow. 

We usually consider $Z_{N}$ strings solutions with the gauge fields
in the CSA with $H_{i},\, i=1,2,\,...,r$ as basis generators which
are the relevant for confinement of the standard monopole solution,
since the monopoles have magnetic flux in direction of the CSA. Then,
as discussed in our previous works \cite{kbrockill2001}\cite{k2003},
since the gauge fields are everywhere in the Cartan subalgebra, from
the BPS equations $D_{z}\phi_{1}=0$ and $D_{\bar{z}}\phi_{1}^{\dagger}=0$
results that the field $\phi_{1}(x)$ is constant and equal to its
asymptotic form (\ref{3.18}), i.e.,\begin{equation}
\phi_{1}(x)=v\cdot H.\label{10}\end{equation}

Then, the BPS equations (\ref{5a})-(\ref{5e}) in the limit $m\rightarrow0$
reduce to\begin{eqnarray}
G_{\bar{z}z} & \rightarrow & -\frac{ie}{4}\left[\phi_{2}^{\dagger},\phi_{2}\right],\nonumber \\
D_{z}\phi_{2} & = & 0,\label{9b}\\
D_{\bar{z}}\phi_{2}^{\dagger} & = & 0,\nonumber \end{eqnarray}
which are exactly the Hitchin's equations \cite{hitchinselfduality}.

As it is known, these equations are equal to a reduction to two dimensions
of the self-duality equation in Euclidean four dimensions \cite{hitchinselfduality},\[
{\cal G}_{\mu\nu}=\frac{1}{2}\epsilon_{\mu\nu\rho\sigma}{\cal G}_{\rho\sigma},\]
 with gauge fields ${\cal W}_{i}=W_{i}$ for $i=1,2$ and \begin{eqnarray}
{\cal W}_{3} & = & \phi_{2r},\label{7.01}\\
{\cal W}_{4} & = & \phi_{2i},\nonumber \end{eqnarray}
where $\phi_{2r}$ and $\phi_{2i}$ are respectively the real and
imaginary parts of $\phi_{2}$ and imposing that the fields does not
depend on the extra dimensions with coordinates $x^{3}$ and $x^{4}$.

The Equations (\ref{9b}) can also be written in the form of a zero
curvature condition  in two dimensions considering the connection
\cite{bonora1} \begin{eqnarray}
{\cal A}_{z} & = & W_{z}+\frac{\lambda}{2}\phi_{2}^{\dagger},\label{7.1}\\
{\cal A}_{\bar{z}} & = & W_{\bar{z}}-\frac{1}{2\lambda}\phi_{2},\nonumber \end{eqnarray}
 where $\lambda$ is a spectral parameter. Then,\begin{eqnarray*}
{\cal F}_{\bar{z}z} & = & \partial_{\bar{z}}{\cal A}_{z}-\partial_{z}{\cal A}_{\bar{z}}+ie\left[{\cal A}_{\bar{z}},{\cal A}_{z}\right]\\
 & = & \left(G_{\bar{z}z}+\frac{ie}{4}\left[\phi_{2}^{\dagger},\phi_{2}\right]\right)+\frac{\lambda}{2}D_{\bar{z}}\phi_{2}^{\dagger}+\frac{1}{2\lambda}D_{z}\phi_{2}.\end{eqnarray*}
Therefore, the system of equations (\ref{9b}) is equivalent to a
zero curvature condition which implies the classical integrability
of the set of  BPS $Z_{N}$ string solutions. More precisely, due
to the limit $m\rightarrow0$, we have the condition ${\cal F}_{\bar{z}z}\rightarrow0$,
which we could call quasi-integrability condition. However, for simplicity
we will use the equal sign in the following integrable equations.

The flat connection ${\cal A}_{z},{\cal A}_{\bar{z}}$ can be written
in terms of the self-dual fields ${\cal W}_{i}$ (\ref{7.01}) as\begin{eqnarray}
{\cal A}_{z} & = & \frac{1}{2}\left({\cal W}_{1}-i{\cal W}_{2}\right)+\frac{\lambda}{2}\left({\cal W}_{3}-i{\cal W}_{4}\right),\label{7.1aa}\\
{\cal A}_{\bar{z}} & = & \frac{1}{2}\left({\cal W}_{1}+i{\cal W}_{2}\right)-\frac{1}{2\lambda}\left({\cal W}_{3}+i{\cal W}_{4}\right).\nonumber \end{eqnarray}

The equivalence of BPS $Z_{N}$ string equations, with the Hitchin's
equations, self-duality equations and zero curvature condition is
interesting because it allows to apply methods and results of these
systems to the $Z_{N}$ string solutions and vice-versa. 

In the next section we show that for $Z_{N}$ string solutions constructed
from different vacuum are associated to different integrable field
equations.

\section{BPS $Z_{N}$ string solutions}

In the Higgs phase of the theory, when $G$ is broken to its center
$C_{G}$ which we consider to be non-trivial, there exist $Z_{N}$
string solutions and the monopoles are confined by these strings.
In order to have finite string tension, asymptotically these solutions
have the form\begin{eqnarray}
\phi_{s}(\varphi,\rho\rightarrow\infty) & = & g(\varphi)\phi_{s}^{\textrm{vac}}g(\varphi)^{-1},\,\,\, s=1,\,2,\label{5.1}\\
W_{i}(\varphi,\rho\rightarrow\infty) & = & \frac{i}{e}\left(\partial_{i}g(\varphi)\right)g(\varphi)^{-1},\,\,\, i=1,2,\nonumber \end{eqnarray}
 where $\phi_{s}^{\textrm{vac}}$ are the vacuum solutions (\ref{3.18}),
(\ref{3.18b}), $\rho$ and $\varphi$ are the radial and angular
coordinates. In order for the configuration to be single valued, $g(\varphi+2\pi)g(\varphi)^{-1}\in C_{G}$.
Considering\[
g(\varphi)=\exp i\varphi M,\,\,\textrm{where}\,\,\, M=\omega\cdot H\]
 it implies that $\exp(2\pi i\omega\cdot H)\in C_{G}$, which results
that $\omega\in\Lambda_{w}(G^{\vee})$, where\begin{equation}
\Lambda_{w}(G^{\vee})=\left\{ \omega=\sum_{i=1}^{r}n_{i}\lambda_{i}^{\vee},\,\,\,\,\,\,\, n_{i}\in\mathbb{Z}\right\} \label{3.3c}\end{equation}
is the coweight lattice of $G$ or equivalently the weight lattice
of the dual group $G^{\vee}$. Then, using the vacuum solutions (\ref{3.18}),
(\ref{3.18b}), the asymptotic form of the $Z_{N}$ string solution
(\ref{5.1}) can be written as \begin{eqnarray}
\phi_{1}(\varphi,\rho\rightarrow\infty) & = & v\cdot H,\nonumber \\
\phi_{2}(\varphi,\rho\rightarrow\infty) & = & \sum_{i=0}^{r}b_{i}\left\{ \exp\left(-i\varphi\omega\cdot\alpha_{i}\right)\right\} E_{-\alpha_{i}},\label{5.4}\\
W_{i}(\varphi,\rho\rightarrow\infty) & = & \frac{\epsilon_{ij}x^{j}}{e\rho^{2}}\omega\cdot H,\,\,\, i=1,2.\nonumber \end{eqnarray}
Therefore, for each weight $\omega$ of the dual group $G^{\vee}$
we can construct a string solution. In \cite{k2003} is shown how
these strings are separated in different topological sectors.

As mentioned before, we can take $\phi_{1}(\varphi,\rho)=v\cdot H$
for the whole space. In order to determine the other fields for the
whole space we consider the Ansatz\begin{equation}
\phi_{2}(\rho,\varphi)=\sum_{i=0}^{r}f_{i}(\rho,\varphi)b_{i}E_{-\alpha_{i}}\exp(-iY(\rho,\varphi)\cdot\alpha_{i}).\label{5.5aa}\end{equation}
 Similarly to the string solution in the Abelian Higgs model, if $\omega$
is such that for a given $\alpha_{i}$, the scalar product $\omega\cdot\alpha_{i}\neq0$,
then the corresponding function $f_{i}(\rho,\varphi)$ must have some
zeros since from the asymptotic form (\ref{5.4}) we see that the
terms with $\omega\cdot\alpha_{i}\neq0$ have non-vanishing winding
number. We can rewrite this Ansatz, similarly to the string solutions
in the Abelian Higgs model \cite{taubes}\cite{manton}, as\begin{equation}
\phi_{2}(\rho,\varphi)=G(\rho,\varphi)\phi_{2}^{\textrm{vac}}G^{-1}(\rho,\varphi)\label{5.5}\end{equation}
where $\phi_{2}^{\textrm{vac}}$ is the vacuum solution (\ref{3.18b})
and\[
G(\rho,\varphi)=\exp\left[Z(\rho,\varphi)\cdot H\right],\,\,\,\,\,\, Z(\rho,\varphi)=-\frac{e}{2}X(\rho,\varphi)+iY(\rho,\varphi),\]
 with $Z(\rho,\varphi),$ $X(\rho,\varphi)$ and $Y(\rho,\varphi)$
being $r$ component real functions with \[
2\ln f_{i}=eX\cdot\alpha_{i}.\]
The points where $f_{i}(\rho,\varphi)$ vanishes, $X\cdot\alpha_{i}$
has a logarithmic singularity. From the asymptotic form (\ref{5.4})
we can conclude that for $\rho\rightarrow\infty$, \begin{eqnarray*}
X(\rho\rightarrow\infty,\varphi) & = & 0,\\
Y(\rho\rightarrow\infty,\varphi) & = & \varphi\omega.\end{eqnarray*}
For the special case of rotationally symmetric solutions, we can consider
$Y(\rho,\varphi)=\varphi\omega$ and $X(\rho,\varphi)=X(\rho)$ is
a radial function. This ansatz can be used for any $Z_{N}$ string
solution, not only BPS.

From Eq. (\ref{5.5}) results that\[
\partial_{z}\phi_{2}=\left[\left(\partial_{z}G\right)G^{-1},\phi_{2}\right].\]
 Therefore, from the BPS equation $D_{z}\phi_{2}=0$, we can conclude
that \begin{equation}
W_{z}=\frac{i}{e}\left(\partial_{z}G\right)G^{-1}+F_{z}=\frac{i}{e}\partial_{z}\left(Z\cdot H\right)+F_{z}\label{5.5a}\end{equation}
where $F_{z}(x)$ is a Lie algebra valued function which commutes
with $\phi_{2}$. On the other hand, from the BPS equation $D_{z}\phi_{1}=0$
and the fact that $\phi_{1}(x)=v\cdot H$, implies $F_{z}(x)$ should
belong to the CSA. Since $\phi_{2}$ is not in the CSA, it implies
that $F_{z}(x)=0$. Similarly, by computing $\partial_{\bar{z}}\phi_{2}^{\dagger}$
and repeating the above argument we can conclude that \begin{equation}
W_{\bar{z}}=\frac{i}{e}\partial_{\bar{z}}\left(G^{-1}\right)^{\dagger}G^{\dagger}=-\frac{i}{e}\partial_{\bar{z}}\left(Z^{\dagger}\cdot H\right).\label{5.5b}\end{equation}
Note that $G^{\dagger}\neq G^{-1}.$ 

Therefore, from the first BPS equation in (\ref{9b}) results that
$X$ satisfies%
\footnote{In order to arrive to this equation we are not considering the points
where $X$ has a singularity. We discuss more on this issue in the
last section.%
}\begin{equation}
\partial_{\bar{z}}\partial_{z}\left(X\cdot H\right)-\frac{e}{4}\left[e^{eX\cdot H}\left(\phi_{2}^{\textrm{vac}}\right)^{\dagger}e^{-eX\cdot H},\phi_{2}^{\textrm{vac}}\right]=0.\label{5.12}\end{equation}
This is the equation of motion of an \textit{Euclidean} two dimensional
integrable system since it equivalent to the zero curvature condition
with the connection (\ref{7.1}) using the fields configurations (\ref{5.5}),
(\ref{5.5a}) and (\ref{5.5b}) (which are solutions of $D_{z}\phi_{2}=0$
and $D_{\bar{z}}\phi_{2}^{\dagger}=0$), that is\begin{eqnarray}
{\cal A}_{z} & = & \frac{i}{e}\partial_{z}(Z\cdot H)+\frac{\lambda}{2}\exp\left(-Z^{\dagger}\cdot H\right)\left(\phi_{2}^{\textrm{vac}}\right)^{\dagger}\exp\left(Z^{\dagger}\cdot H\right),\label{5.12a}\\
{\cal A}_{\bar{z}} & = & -\frac{i}{e}\partial_{\bar{z}}(Z^{\dagger}\cdot H)-\frac{1}{2\lambda}\exp\left(Z\cdot H\right)\phi_{2}^{\textrm{vac}}\exp\left(-Z\cdot H\right).\nonumber \end{eqnarray}
 Using the fact that $\phi_{2}^{\textrm{vac}}=\sum_{l=0}^{r}b_{l}E_{-\alpha_{l}},$
we can write Eq. (\ref{5.12}) as\begin{equation}
\partial_{\bar{z}}\partial_{z}X-\frac{e}{4}\sum_{j=0}^{r}b_{j}^{2}\alpha_{j}^{\vee}e^{e\alpha_{j}\cdot X}=0,\label{5.13}\end{equation}
remembering that $X$ is an $r$ component scalar field. For the vacuum
solutions with $b_{0}=0$, we define\[
X_{\alpha_{i}}=\alpha_{i}\cdot X,\,\,\,\,\,\,\, i=1,\,2,\,...,\, r.\]
Then, Eq. (\ref{5.13}) can be written as\begin{equation}
\partial_{\bar{z}}\partial_{z}X_{\alpha_{i}}-\frac{e}{4}\sum_{j=1}^{r}K_{ij}b_{j}^{2}\exp\left(eX_{\alpha_{j}}\right)=0,\label{5.14a}\end{equation}
where $K_{ij}$ is the Cartan matrix (\ref{3.4}). 

On the other hand, for vacuum solutions with $b_{0}\neq0$, we define\begin{equation}
X_{\alpha_{i}}=\alpha_{i}\cdot X,\,\,\,\,\,\, i=0,\,1,\,2,\,...,\, r.\label{5.14aa}\end{equation}
In this case, Eq. (\ref{5.13}) can be written as\begin{equation}
\partial_{\bar{z}}\partial_{z}X_{\alpha_{i}}-\frac{e}{4}\sum_{j=0}^{r}\widehat{K}_{ij}b_{j}^{2}\exp\left(eX_{\alpha_{j}}\right)=0,\,\,\,\,\,\, i=0,\,1,\,2,\,...,\, r\label{5.14b}\end{equation}
where $\widehat{K}_{ij}$ is the extended Cartan matrix which has
the same form as the Cartan matrix Eq. (\ref{3.4}), but with $i,j=0,1,\,...,r$,
and are associated to the untwisted affine Lie algebras. However,
from (\ref{3.5}) and the fact that $\alpha_{0}=-\psi$ we can conclude
that \[
\sum_{i=0}^{r}m_{i}\alpha_{i}^{\vee}=0,\]
where we consider $m_{0}=1$. Therefore, the fields $X_{\alpha_{i}}$
in (\ref{5.14aa}) are not independent but satisfy the constraint
\[
\sum_{i=0}^{r}\frac{2m_{i}}{\alpha_{i}^{2}}X_{\alpha_{i}}=0.\]
 Therefore, equation (\ref{5.14b}) must be subject to this constraint
\cite{kneippolivetwist}. 

As we mentioned before, for a $Z_{N}$ string solution associated
to a vector $\omega$, for the terms in Eq. (\ref{5.5aa}) where $\omega\cdot\alpha_{i}\neq0$,
the corresponding function $f_{i}$ must have some zeros and hence
$X_{\alpha_{i}}=\alpha_{i}\cdot X$ has logarithmic singularities.
Therefore equations (\ref{5.14a}) and (\ref{5.14b}) are valid except
at the singularities of $X_{\alpha_{i}}$. Similarly to the Abelian
case \cite{taubes}\cite{rebbi}, we can allow for these singularities
by including delta-functions on the right hand side of the above equations.

\section{A vacuum solution and affine Toda field theories }

Let us now consider a concrete vacuum solution. In order to be a vacuum
solution of the potential (\ref{8}), the constants in (\ref{3.18}),
(\ref{3.18b}), must satisfy the relation\begin{equation}
m\left(K^{-1}\right)_{ij}v_{j}=b_{i}^{2}-m_{i}b_{0}^{2}.\label{4.1}\end{equation}
 In \cite{k2003}\cite{kneippsine} we analyzed two vacuum solutions
with $b_{0}=0$, which are valid for any gauge group $G$: 

a) The first vacuum solution we considered was \begin{eqnarray}
v_{i} & = & a,\label{4.2}\\
b_{i} & = & \sqrt{ma\sum_{j=1}^{r}\left(K^{-1}\right)_{ij}}=\sqrt{ma\delta\cdot\lambda_{i}}\,,\,\,\,\,\,\, i=1,2,\,...,\, r,\end{eqnarray}
where $a$ is a positive real constant and $\delta=\sum_{i=1}^{r}\lambda_{i}^{\vee}$
is the dual Weyl vector. With this vacuum, the $Z_{N}$ strings tensions
satisfy the Casimir scaling \cite{k2003}. 

b) The second vacuum solution was\begin{eqnarray}
v_{i} & = & ay_{i}^{(1)},\label{4.4}\\
b_{i} & = & \frac{1}{2\sin\frac{\pi}{2h}}\sqrt{may^{(1)}}\,,\,\,\,\,\,\, i=1,2,\,...,\, r,\end{eqnarray}
where $a$ is a positive real constant and $y_{i}^{(1)}$are the components
of the Perron-Frobenius eigenvectors of $K_{ij}$ associated to the
eigenvalue $4\sin^{2}\frac{\pi}{2h}$. With this vacuum, the $Z_{N}$
strings tensions satisfy the sine law scaling \cite{kneippsine}. 

As mentioned before, in order for the $Z_{N}$ string's BPS equations
to be consistent with the equations of motion we must take the limit
$m\rightarrow0$. Therefore, for the constants $b_{i}$ or equivalently
$\phi_{2}^{\textrm{vac}}$ to be finite, we must take $a\rightarrow\infty$
(and therefore $v_{i}\rightarrow\infty)$ keeping the product $ma$
finite \cite{kneippsine}. Another possibility we use here is to consider
the vacuum solutions (\ref{3.18}), (\ref{3.18b}) with $b_{0}\neq0$
in which case we can keep $a$ finite. One can see this from Eq. (\ref{4.1})
which implies that \begin{equation}
b_{i}=\sqrt{m\left(K^{-1}\right)_{ij}v_{j}+m_{i}b_{0}^{2}}.\label{4.6}\end{equation}
From this equation, considering that $v_{j}$ is finite, when we take
$m\rightarrow0$, it implies that\[
b_{i}\rightarrow\sqrt{m_{i}}b_{0}\]
which is finite with $b_{0}$ being an arbitrary non-vanishing constant.
That result hold when the components $v_{j}$ satisfy either (\ref{4.2})
or (\ref{4.4}). In each case, the ratio of tensions of the BPS $Z_{N}$
strings will continue to satisfy the Casimir scaling or the sine law
scaling respectively. The vacuum solution (\ref{3.18b}) will stay\begin{equation}
\phi_{2}^{\textrm{vac}}=b_{0}\sum_{l=0}^{r}\sqrt{m_{i}}E_{-\alpha_{l}}=b_{0}E,\label{4.7}\end{equation}
where\[
E=\sum_{i=0}^{r}\sqrt{m_{i}}E_{\alpha_{i}}\]
and therefore asymptotically\begin{equation}
\phi_{2}(\varphi,\rho\rightarrow\infty)=b_{0}\sum_{i=0}^{r}\sqrt{m}_{i}\left\{ \exp\left(-i\varphi\omega\cdot\alpha_{i}\right)\right\} E_{-\alpha_{i}}.\label{4.7a}\end{equation}
The generator $E$ satisfy $[E,E^{\dagger}]=0$. Hence it is diagonalizable
and can be embedded in a new Cartan subalgebra. This generator was
originally introduced by Konstant \cite{konstant}.

For this vacuum we can write (\ref{5.12}) as\begin{equation}
\partial_{\bar{z}}\partial_{z}\left(X\cdot H\right)-\frac{eb_{0}^{2}}{4}\left[e^{eX\cdot H}E^{\dagger}e^{-eX\cdot H},E\right]=0,\label{4.8}\end{equation}
 It can also be written as \begin{equation}
\partial_{\bar{z}}\partial_{z}X-\frac{eb_{0}^{2}}{4}\sum_{j=0}^{r}m_{j}\alpha_{j}^{\vee}\exp\left(e\alpha_{j}\cdot X\right)=0\label{4.10}\end{equation}
or\begin{equation}
\partial_{\bar{z}}\partial_{z}X_{\alpha_{i}}-\frac{eb_{0}^{2}}{4}\sum_{j=0}^{r}\widehat{K}_{ij}m_{j}\exp\left(eX_{\alpha_{j}}\right)=0.\label{4.10a}\end{equation}

Eq. (\ref{4.8}) (or (\ref{4.10}), (\ref{4.10a}) ) is the equation
of motion of Euclidean two dimensional integrable affine Toda field
theory associated to the untwisted affine Lie algebra $\widehat{g}$
obtained from $g$, with coupling constant $e$, equal to the coupling
constant of the gauge theory, and mass parameter equal to $eb_{0}$.
In \cite{kneippsine} was shown that the spectrum of BPS string tensions
in the second vacuum solution (\ref{4.4}) coincide with the solitons
mass spectrum of the corresponding affine Toda theory.

It is interesting to note that the monopole's BPS equations with spherical
symmetry reduces to the equation of conformal Toda theory \cite{GanoulisGodOlive}.
The equation of Affine Toda theory was also obtained from Hitchin's
equations for $U(N)$ matrix theory in \cite{bonora1}, but with some
differences as we discuss in the next section. A relation between
Hitchin equations and integrable systems was also consider in \cite{Hitchin-Ward}.
For $g=su(2)$, considering $\alpha_{1}=1=-\alpha_{0}$, equation
(\ref{4.10}) reduces to the sinh-Gordon equation\[
\partial_{\bar{z}}\partial_{z}X-\frac{eb_{0}^{2}}{2}\sinh\left(eX\right)=0.\]

\section{Rotationally symmetric solutions}

Let us now consider the special case of rotationally symmetric solutions.
In this case for a $Z_{N}$ string associated with the vector $\omega$
of the weight lattice of dual group $G^{\vee}$, $Y(\rho,\varphi)=\varphi\omega$
and $X(\rho,\varphi)=X(\rho)$ is a radial function, and hence\begin{equation}
Z(\rho,\varphi)=-\frac{e}{2}X(\rho)+i\varphi\omega.\label{5.20}\end{equation}
Therefore, in this case the scalar fields has the form \begin{eqnarray}
\phi_{1}(\varphi,\rho) & = & v\cdot H,\label{5.20a}\\
\phi_{2}(\varphi,\rho) & = & b_{0}\sum_{i=0}^{r}\sqrt{m_{i}}\left\{ \exp\left(\frac{e}{2}X\cdot\alpha_{i}-i\varphi\omega\cdot\alpha_{i}\right)\right\} E_{-\alpha_{i}}.\label{5.20b}\end{eqnarray}
Generalizing some results from stringy instantons of matrix string
theories \cite{bonora1}\cite{matrix}, we can determine the form
of the gauge fields: since $2\varphi=-i\ln(z/\bar{z})$, the gauge
fields (\ref{5.5a}) and (\ref{5.5b}) are given by\begin{eqnarray}
W_{z} & = & \frac{i}{2}\left[-\partial_{z}\left(X\cdot H\right)+\frac{1}{ez}\omega\cdot H\right],\label{5.21}\\
W_{\bar{z}} & = & -\frac{i}{2}\left[-\partial_{\bar{z}}\left(X\cdot H\right)+\frac{1}{e\bar{z}}\omega\cdot H\right].\nonumber \end{eqnarray}
 Using that $\partial_{z}(1/\bar{z})=\pi\delta^{(2)}(x)$, we obtain
that the magnetic field of the $Z_{N}$ string is\begin{equation}
B_{3}=-2iG_{\bar{z}z}=-2\left[\partial_{\bar{z}}\partial_{z}\left(X\cdot H\right)-\frac{\pi}{e}\omega\cdot H\delta^{(2)}(x)\right].\label{5.21a}\end{equation}
From the requirement of regularity at $z=0$ implies that near the
origin\[
\partial_{\bar{z}}\partial_{z}\left(X\cdot H\right)\sim\frac{\pi}{e}\omega\cdot H\delta^{(2)}(x)+\textrm{const}.\]
Therefore, near the origin\begin{equation}
X(\rho\rightarrow0)\sim-\frac{2}{e}\omega\ln\left|z\right|+\textrm{const.}\label{4.20}\end{equation}
 On the other hand, for $\rho\rightarrow\infty$, $X(\rho\rightarrow\infty)\rightarrow0$,
as we mentioned before. Eq. (\ref{4.20}) is consistent with the fact
that in the general case (not necessarily rotationally symmetric),
$X_{\alpha_{i}}=\alpha_{i}\cdot X$ has a logarithmic singularity
if $\alpha_{i}\cdot\omega\neq0$. 

From (\ref{5.20a}) and (\ref{5.21a}) we conclude that the flux (\ref{7b})
of a string associated to the coweight $w$ is \[
\Phi_{\textrm{st}}=\frac{2\pi}{e}\frac{\omega\cdot v}{\left|v\right|},\]
which is consistent with the result in \cite{k2003} using just the
asymptotic form of the $Z_{N}$ string solution (\ref{5.4}).

From Eq. (\ref{4.10a}) and the behavior of the solution near the
origin (\ref{4.20}) we can conclude that for the special case of
rotationally symmetric solutions, for a $Z_{N}$ string associated
with a coweight $w$, the radial function $X(\rho)$ must satisfy
\begin{equation}
\frac{\partial^{2}X}{\partial\rho^{2}}+\frac{1}{\rho}\frac{\partial X}{\partial\rho}-eb_{0}^{2}\sum_{j=0}^{r}\alpha_{j}^{\vee}m_{j}\exp\left(e\alpha_{j}\cdot X\right)=\frac{4\pi}{e}\omega\delta^{(2)}(x)\label{4.12a}\end{equation}
or\begin{equation}
\frac{\partial^{2}X_{\alpha_{i}}}{\partial\rho^{2}}+\frac{1}{\rho}\frac{\partial X_{\alpha_{i}}}{\partial\rho}-eb_{0}^{2}\sum_{j=0}^{r}\widehat{K}_{ij}m_{j}\exp\left(eX_{\alpha_{j}}\right)=\frac{4\pi}{e}\omega\cdot\alpha_{i}\delta^{(2)}(x).\label{4.12}\end{equation}
 One could arrive directly to this equation with the singularity using
(\ref{5.20}) and (\ref{5.21}) in the BPS equation (\ref{5a}) or
in the connection (\ref{5.12a}). That result is similar to the string
solution in the Abelian-Higgs theory where for a rotationally symmetric
configuration Ansatz the radial function, satisfies a rotationally
symmetric form of a Liouville's equation plus a constant with a $\delta$-function
at the origin\cite{taubes}\cite{rebbi}. Note that our condition
$m\rightarrow0$ corresponds to take the limit $a\rightarrow0$ for
the constant appearing in the potential in Abelian-Higgs theory, in
which case the radial function of the string solution would satisfy
Liouville's equation with a $\delta$-function at the origin.

The $Z_{N}$ string solutions have great similarity with the Riemannian
or stringy instantons of matrix string theories \cite{bonora1}\cite{matrix},
but also some differences: in the matrix stringy theories there is
no gauge symmetry breaking. For the stringy instantons, the scalar
field has a branch cut in the origin instead of a zero for $\phi_{2}(\rho,\varphi)$
for the rotationally symmetric $Z_{N}$ strings. They also have different
angular dependence since the $Z_{N}$ string solutions are associated
to classes of $\Pi_{1}(G/C_{G})$. Moreover, the affine Toda equation
associated to stringy instantons \cite{bonora1} has a singularity
structure different from (\ref{4.12a}).

Eqs. (\ref{4.12a}) or (\ref{4.12}) with source is equivalent to
the homogeneous equation with the boundary (\ref{4.20}) near the
origin. In general to solve {}``Toda type theories'' one can apply
Leznov-Saveliev method \cite{Saveliev}. Solutions for a similar equation
for $g=su(2)$ with singularity were analyzed in \cite{luizbonora}.
Soliton solutions of Affine Toda theories (which have different boundary
condition) were analyzed for example in \cite{solitonsATFT}\cite{kneippolivetwist}
for different algebras $g$. Similar methods may be can applied for
this case with different boundary conditions.

\end{document}